\documentclass[a4paper,
               keeplastbox,   
               ]{jacow}
\usepackage{pdfpages,multirow,ragged2e} %
\makeatletter%
	\ifboolexpr{bool{xetex}}
	 {\renewcommand{\Gin@extensions}{.pdf,%
	                    .png,.jpg,.bmp,.pict,.tif,.psd,.mac,.sga,.tga,.gif,%
	                    .eps,.ps,%
	                    }}{}
\makeatother

\ifboolexpr{bool{xetex} or bool{luatex}} 
 {}                                      
 {\usepackage[utf8]{inputenc}}           

\usepackage[USenglish]{babel}

\begin{document}
\title{A Realistic Proportional-Integral RF Feedback Model for Longitudinal Beam Dynamics Simulation}

\author{Tianlong He\thanks{Contact author: htlong@ustc.edu.cn}, Wenshu Liang, Jincheng Xiao, Xin Huang
\\
	National Synchrotron Radiation Laboratory, USTC, Hefei 230029, China}
\maketitle

\begin{abstract}
Modern fourth-generation storage ring light sources predominantly utilize digital I/Q-based proportional-integral (PI) feedback for their radio-frequency (RF) systems. This paper introduces a dedicated PI feedback model implemented in the STABLE tracking code to enable accurate and fast longitudinal beam dynamics simulations. The model's key innovation lies in its treatment of the continuous generator current, which is discretized into electron-bunch-like charge pulses, while the cavity voltage is refreshed on an RF-cycle basis. This methodology offers a more physically accurate model of the beam-cavity-feedback coupling, providing a versatile tool for precise longitudinal beam dynamics studies in single- and multi-rf configurations.
\end{abstract}

\section{INTRODUCTION}
Achieving precise regulation of radio-frequency (RF) cavities is fundamental to the performance of modern storage ring light sources, a task predominantly accomplished by digital I/Q-based proportional-integral (PI) feedback systems. Accurate simulation of the coupled beam-cavity-feedback dynamics is therefore essential, especially for applications like bunch lengthening with harmonic cavities. Existing multi-particle tracking codes provide various approaches to this problem. The ELEGANT code, for example, incorporates an RF feedback model originally designed for analog amplitude/phase systems~\cite{berenc2015modeling}, requiring the conversion of PI parameters into equivalent filter coefficients for simulation~\cite{Wang2023RFFeedback}. Other codes, such as MBTRACK2, directly implement a PI controller framework to model the cavity response~\cite{mbtrack2_ipac21,Yamamoto2018}. While these tools accurately model fundamental cavity dynamics, parameter scans of modern digital feedback systems—including PI gains, loop delays, and operating conditions—to evaluate their impact on beam behavior demand considerable computational effort.

This paper presents a refined, physics-based RF feedback model implemented within the GPU-accelerated STABLE tracking code~\cite{STABLE_HTL}. The model is natively designed for digital I/Q PI feedback systems and introduces two key advances. First, it employs a high-fidelity cavity model where the continuous generator current is discretized into equivalent charge pulses, and thus the cavity voltage is updated iteratively on an RF-cycle basis, ensuring an accurate representation of transient beam-cavity-feedback coupling interactions. Second, and distinctively, STABLE leverages NVIDIA CUDA for massive parallelization on GPUs. Compared to CPU-parallelized frameworks, this architecture provides a substantial increase in computational throughput for large-scale parameter scans at a lower hardware cost. This combination of physical fidelity and computational power makes STABLE a practical and efficient tool for investigating longitudinal beam instability in the context of modern feedback systems and challenging scenarios like bunch-lengthening harmonic cavity operation.

\section{DISCRETIZATION OF THE GENERATOR CURRENT}
\begin{figure}[h]
    \centering
    \includegraphics[width=0.45\textwidth]{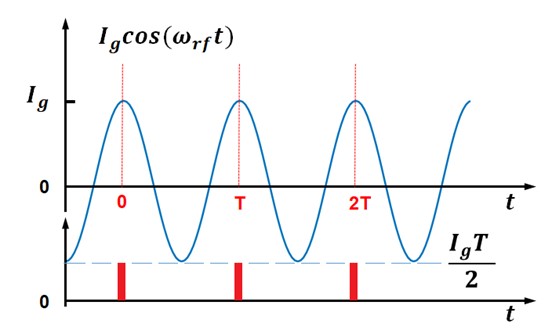}
    \caption{Discretization of a cosine generator current into equivalent pulsed charges.}
    \label{fig1}
\end{figure}

To facilitate the numerical calculation of the generator voltage phasor, it is advantageous to discretize the continuous cosine generator current into pulsed charges with the same period. As illustrated in Fig.~\ref{fig1}, we demonstrate that the charge quantity of each pulse equals $\tfrac{1}{2} I_g T$, where $I_g$ and $T$ are the amplitude and period of the generator current, respectively.

Analogous to representing a longitudinally distributed bunch charge by an equivalent point charge with a form factor, the equivalent pulse charge for a cosine generator current can be obtained by:
\begin{equation}\label{eq:1}
  Q_g = \int_{-T/2}^{T/2} I_g \cos(\omega_{\text{rf}} t) e^{-j\omega_r t}\, dt
      = \frac{1}{2} I_g T ,
  \tag{1}
\end{equation}
where the approximation $\omega_r \approx \omega_{\text{rf}}$ is used to compute the integral.

It is convenient to represent RF signals (voltages and currents) as complex phasors. In a counterclockwise rotating reference frame with frequency $\omega_{\text{rf}}$, we denote the generator current phasor and the equivalent pulse charge phasor as $\tilde{I}_g$ and $\tilde{Q}_g$, respectively. For a pulsed charge excitation at time zero, the resulting cavity voltage phasor is:
\begin{equation}\label{eq:2}
  \tilde{V}_g = \frac{\tilde{Q}_g \omega_r R_L}{Q_L} .
  \tag{2}
\end{equation}
Here, $\omega_r$, $R_L$, and $Q_L$ are the angular resonant frequency, loaded shunt impedance, and loaded quality factor of the cavity, respectively. Substituting Eq.~\eqref{eq:1} into Eq.~\eqref{eq:2} and using $\omega_r T \approx 2\pi$ yields:
\begin{equation}\label{eq:3}
  \tilde{V}_g = \tilde{I}_g \frac{\pi R_L}{Q_L} .
  \tag{3}
\end{equation}
Using Eq.~\eqref{eq:3}, the cavity voltage phasor induced by the generator current can be computed iteratively:
\begin{equation}\label{eq:4}
  \tilde{V}_g^{\,n}
  = \tilde{V}_g^{\,n-1} e^{\left(j-\frac{1}{2Q_L}\right)\omega_r T}
    + \tilde{I}_g^{\,n} \frac{\pi R_L}{Q_L} ,
  \tag{4}
\end{equation}
where the superscript $n$ denotes the iteration index.

Equation~\eqref{eq:4} highlights that the change in cavity voltage due to \textbf{a variation in the generator current is not instantaneous but occurs over a finite time}. This timescale is characterized by the cavity time constant $\tau = 2 Q_L / \omega_r$. For typical superconducting cavities with $Q_L \sim 10^5$ and $\omega_r \sim 2\pi \times 500~\text{MHz}$, $\tau \sim 64~\mu\text{s}$, which corresponds to approximately 40 beam revolutions in the HALF storage ring. Consequently, \textbf{it is physically inaccurate to use a turn-averaged cavity voltage to directly regulate the generator voltage on a per-turn basis}. A more realistic approach is to adjust the generator current phasor $\tilde{I}_g$ via a PI controller, as commonly implemented in low-level RF feedback systems.

For implementation in a macro-particle tracking code, the cavity voltage phasor is updated bucket by bucket (i.e., every RF period) using Eq.~\eqref{eq:4}. The remaining task is to determine the generator current phasor $\tilde{I}_g$ at each step, which is governed by the feedback model described in the following section.

\section{A REALISTIC RF FEEDBACK MODEL}
The digital low-level radio frequency (DLLRF) feedback system employing I/Q control is standard in modern storage ring light sources. In this scheme, RF signals are decomposed into in-phase (I) and quadrature (Q) components. In our tracking simulations, we represent these signals using complex phasors, where the I and Q components correspond directly to the real and imaginary parts. To accurately model LLRF feedback for beam dynamics simulations, it is essential to follow a realistic signal processing chain. This work references the LLRF system at the SSRF storage ring~\cite{xia2019transfer}, which comprises four control loops: amplitude, phase, tuning, and direct feedback. The amplitude and phase loops, jointly termed the field loop, regulate and stabilize the cavity field. Given their relatively long response time compared to the field loop, and since the direct feedback loop primarily mitigates heavy beam loading, both are neglected in our tracking simulations for simplicity. Thus, we focus solely on modeling the field loop. We also assume ideal signal processing, ignoring RF noise and nonlinearities.

\begin{figure}[h]
    \centering
    \includegraphics[width=0.45\textwidth]{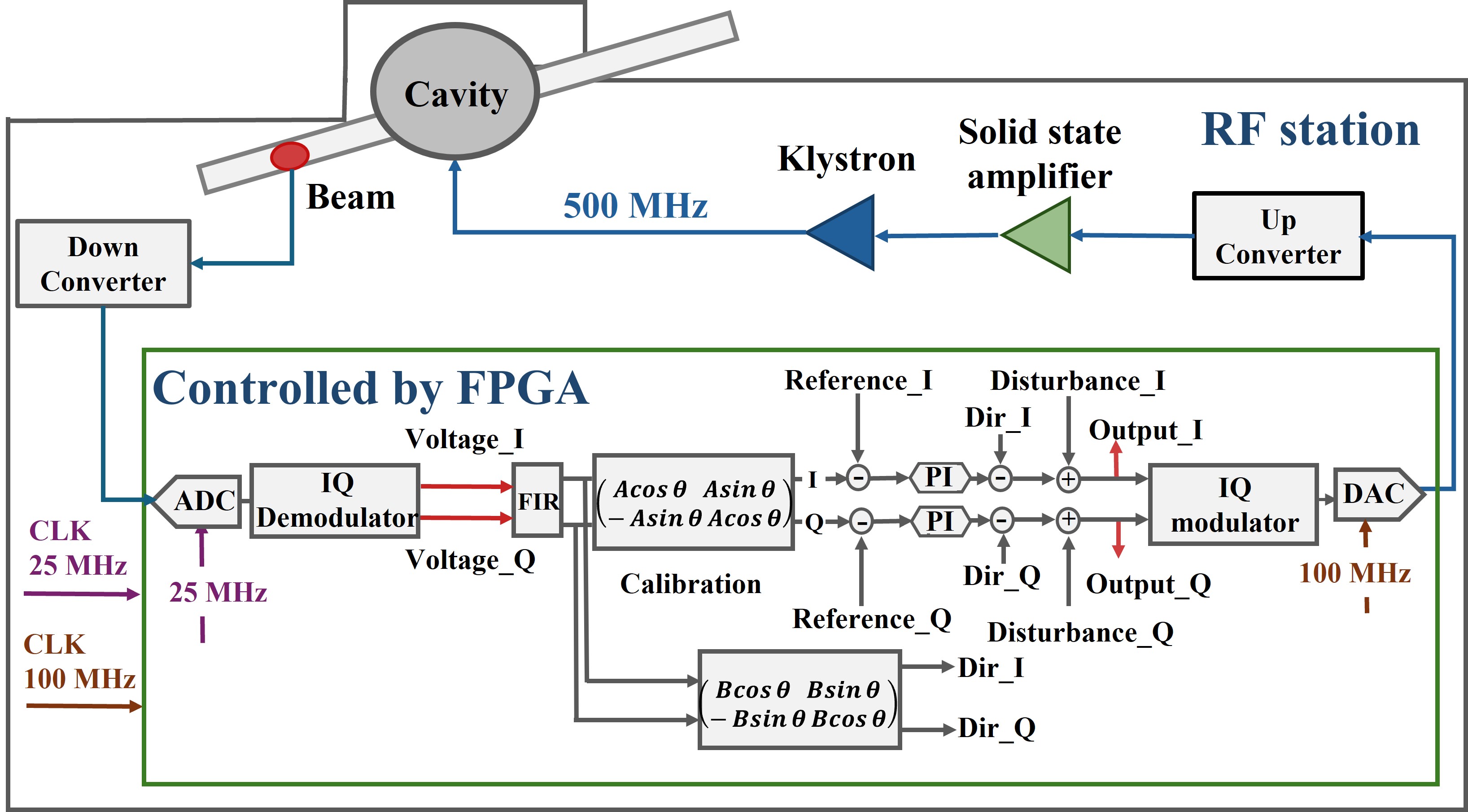}
    \caption{Simplified block diagram of the RF system at SSRF~\cite{xia2019transfer}.}
    \label{fig2}
\end{figure}

The field loop operation, detailed in~\cite{xia2019transfer}, involves several stages. The cavity voltage signal (e.g., 500 MHz) is picked up, down-converted to an intermediate frequency (IF, e.g., 31.25 MHz), and then sampled by an Analog-to-Digital Converter (ADC) at a clock frequency (e.g., 25 MHz). The ADC samples are transformed into I/Q components using a non-I/Q demodulation algorithm (with a ratio such as $f_{\text{IF}} / f_{\text{clock}} = 5/4$). These I/Q components are filtered by a Finite Impulse Response (FIR) low-pass filter, often implemented as a 64-tap (or less tap) averaging algorithm. The filtered I/Q data is calibrated (e.g., via a rotation module for loop stability) and then compared to reference setpoints to generate I/Q error signals. These errors are fed into two PI controllers (which, due to path symmetry, can have identical proportional and integral gains) to produce corrected I/Q values. Finally, these corrected I/Q signals are modulated back to an IF digital signal, converted to an analog signal by a Digital-to-Analog Converter (DAC), up-converted to the RF frequency, amplified, and used to drive the cavity.

\begin{figure}[h]
    \centering
    \includegraphics[width=0.45\textwidth]{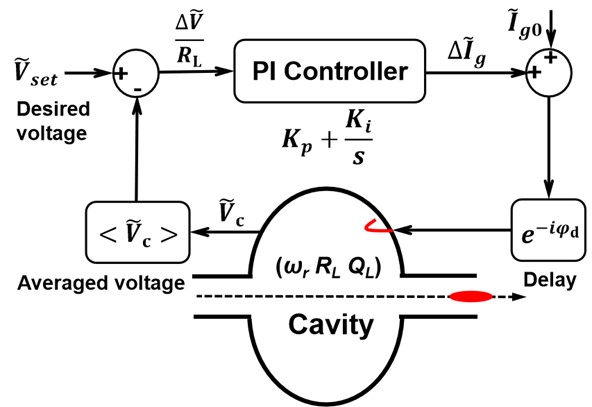}
    \caption{Schematic of the modeled LLRF control system for simulation.}
    \label{fig3}
\end{figure}

Based on this workflow, we propose a simplified yet realistic model for the field loop, depicted in Fig.~\ref{fig3}. The feedback procedure in our simulation is divided into three sequential stages.

\subsection{Stage 1: Before the PI Controller}
The 64-tap FIR filter operation is equivalent to averaging the cavity voltage phasor over a specific time window. This averaging period is determined by the RF frequency, the IF, the ADC sampling rate, and the demodulation algorithm. For the 500 MHz RF system, this corresponds to 5120 RF periods (buckets). It should be noted that for RF systems employing a higher sampling rate or a smaller number of FIR taps, the required number of buckets for computing the averaged voltage would be significantly fewer. In the simulation, the cavity voltage phasor is updated every bucket, making it straightforward to compute a moving average over the last N buckets. In practice, this parameter N can be estimated based on the specific LLRF system configuration and directly set within the STABLE code accordingly. The voltage error phasor $\Delta \tilde{V}$ is generated by subtracting this averaged voltage phasor from the reference setpoint phasor $\tilde{V}_{\text{set}}$. For simplicity, this voltage error is converted into an equivalent current error $\Delta \tilde{V}/R_L$ before being sent to the PI controller, where $R_L$ is the loaded shunt impedance.

\subsection{Stage 2: During the PI Controller}
The PI controller receives the current error $e^n=\left(\Delta \tilde{V}/R_L\right)^n$ and outputs a correction to the generator current phasor, $(\Delta \tilde{I}_g)^n$. The controller transfer function in the Laplace domain is
$K_p+K_{i,\mathrm{cont}}/s$, where $K_p$ and $K_{i,\mathrm{cont}}$ denote the proportional gain and the \emph{continuous-time} integral gain, respectively. In the discrete-time simulation, the PI controller is updated every $T_s$ and the integral term is implemented by a forward-Euler accumulator:
\begin{equation}
  I^n = I^{n-1} + K_{i,\mathrm{cont}}\,T_s\, e^n ,
  \label{eq:pi_integrator_update}
  \tag{5}
\end{equation}
and 
\begin{equation}
  (\Delta \tilde{I}_g)^n = K_p\, e^n + I^n .
  \label{eq:pi_output_update}
  \tag{6}
\end{equation}
Equivalently, by defining the \emph{discrete-time} integral gain $K_I \triangleq K_{i,\mathrm{cont}}T_s$, the above equation can be written as
\begin{equation}
  (\Delta \tilde{I}_g)^n
  = K_p \left(\frac{\Delta \tilde{V}}{R_L}\right)^n
    + K_I \sum_{i=1}^{n} \left(\frac{\Delta \tilde{V}}{R_L}\right)^i .
  \tag{7}
\end{equation}
Here, $n$ is the controller update index, while $K_p$ and $K_I$ are the discrete-time proportional and integral gains used in the tracking simulation.
\subsection{Stage 3: After the PI Controller}
Following the approach in Ref.~\cite{berenc2015modeling}, the correction $\Delta \tilde{I}_g$ is added to a fixed, initial generator current $\tilde{I}_{g0}$. This initial current facilitates setting the operating point or incorporating feed-forward, and it can be determined from steady-state conditions:
\begin{equation}
  \tilde{I}_{g0} R_L \cos(\psi)\, e^{-j\psi}
  = \tilde{V}_{g0}
  = \tilde{V}_{\text{set}} - \langle \tilde{V}_{b0} \rangle,
  \tag{8}
\end{equation}
where $\psi$ is the cavity detuning angle, $\tilde{V}_{\text{set}}$ is the desired cavity voltage, and $\langle \tilde{V}_{b0} \rangle$ is the initial average beam-induced voltage phasor across all buckets.

The total generator current phasor, $\tilde{I}_g = \Delta \tilde{I}_g + \tilde{I}_{g0}$, is then subjected to a delay before driving the cavity model (using Eq.~\eqref{eq:4}). This delay accounts for the finite time required for signal transmission and processing in a real system, typically on the order of microseconds. For the HALF ring example, we assume a delay of $1.6~\mu\text{s}$, corresponding to approximately 800 RF buckets.

\section{Summary}
The presented model provides a practical tool for studying the influence of key parameters---such as the PI gains $K_p$, $K_i$, and the feedback loop delay---on longitudinal beam dynamics. Such studies are essential for fourth-generation light sources, where significant bunch lengthening (by factors of larger than 4) via harmonic cavities is desired. This lengthening can potentially introduce instabilities that challenge the performance of the main RF cavity's DLLRF system. 

This model has been successfully applied to simulate and analyze how the main RF cavity's PI feedback influences instabilities driven by the fundamental mode of a passive superconducting harmonic cavity, demonstrating its direct relevance to beam stability in double RF configurations~\cite{STABLE_HTL2}. 

Furthermore, the model's structure allows for straightforward extension to investigate other effects. For instance, incorporating a modulation module to perturb the generator current would enable the simulation of RF noise impact on beam dynamics. This extension, however, is beyond the scope of the present paper and is planned for future work.

\section{ACKNOWLEDGEMENTS}
This work was supported by the Fundamental Research Funds for the Central Universities (No. WK2310000127) and the National Natural Science Foundation of China (Grants No. 12375324 and No. 12105284).

\bibliographystyle{unsrt}
\bibliography{refs}

\end{document}